# The nonlinear dynamics of a cantilever beam subject to axial flow in a tapered passage

Filipe Soares[1,2,a)], José Antunes[2,3], Christophe Vergez[1],
Vincent Debut[2,3], Bruno Cochelin[1], Fabrice Silva[1]

[1] *Aix-Marseille Univ, CNRS, Centrale Marseille, LMA UMR7031, Marseille, France.*
[2] *Instituto Superior Técnico, Centro de Ciências e Tecnologias Nucleares, Lisbon, Portugal.*
[3] *Instituto de Etnomusicologia, Centro de Estudos em Música e Dança, UNL, Lisbon, Portugal.*
[a)] *Corresponding author: filipe.soares@ctn.tecnico.ulisboa.pt*

**Abstract.** A cantilever beam under axial flow, confined or not, is known to develop self-sustained oscillations at sufficiently large flow velocities. In recent decades, the analysis of this archetypal system has been mostly pursued under linearized conditions, to calculate the critical boundaries separating stable from unstable behavior. However, nonlinear analysis of the self-sustained oscillations ensuing flutter instabilities are considerably rarer. Here we present a simplified one-dimensional nonlinear model describing a cantilever beam subjected to confined axial flow, for generic axial profiles of the fluid channels. In particular, we explore how the shape of the confinement walls affects the dynamics of the system. To simplify the problem, we consider symmetric channels with plane walls in either converging or diverging configurations. The beam is modeled in a modal framework, while bulk-flow equations, including singular head-loss terms, are used to model the flow-structure coupling forces. The dynamics of the system are first analyzed through linear stability analysis to assess the stabilizing/destabilizing effects of the channel walls configuration. Subsequently, we develop a systematic nonlinear analysis based on the continuation of periodic solutions. The harmonic balance method is used in conjunction with the asymptotic numerical method to calculate branches of periodic solutions. The continuation-based methods are used to investigate bifurcations with respect to both the reduced flow velocity and the channel slope parameter (expanding or narrowing). From the results presented, we illustrate how continuation-based approaches and bifurcation analysis provide an efficient tool to analyze the nonlinear behavior of flow-induced vibration problems, particularly when reduced/simplified models are available.

## MODEL DESCRIPTION

The model studied here describes the fluid-structure interaction of a cantilevered beam in symmetric confined flow with expanding/narrowing channel walls, as illustrated in Figure 1.

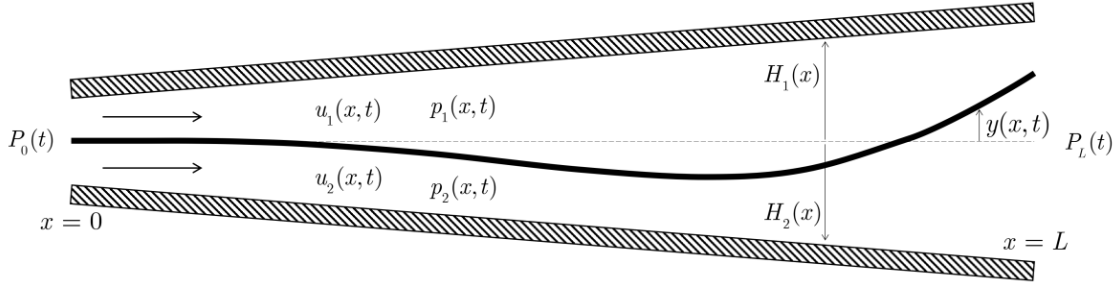

**FIGURE 1**. Diagram of the considered 1D model.

The dynamics of the linear cantilevered beam are defined in a modal framework. The vertical displacement of the beam $y(x,t)$ is developed in terms of $M$ modes and the resulting set of ODEs read

$$\ddot{q}_m(t) + 2\omega_m \zeta_m \dot{q}_m(t) + \omega_m^2 q_m(t) = F_m(t)/m_m \quad \text{with} \quad y(x,t) = \sum_{m=1}^{M} \varphi_m(x) q_m(t) \qquad (1)$$

where $q_m(t)$, $\varphi_m(x)$, $m_m$, $\omega_m$ and $\zeta_m$ are the modal displacements, shapes, masses, frequencies and damping ratios, respectively. The external modal forces are given by the projection of the flow pressure fields (upper and lower sides of the beam) unto the modal basis

$$F_m(t) = M^* \int_0^1 [p_2(x,t) - p_1(x,t)] \varphi_m(x) dx \qquad (2)$$

where the fluid-beam mass ratio is written explicitly as $M^* = \rho L / \rho_s e$, with $\rho$ and $\rho_s$ the densities of the fluid and the beam, while $e$ denotes the thickness of the beam. Assuming incompressible and inviscid flow, the momentum and continuity equations for the flow in each channel $c$ are given in dimensionless form by

$$\frac{\partial u_c}{\partial t} + u_c \frac{\partial u_c}{\partial x} + \frac{\partial p_c}{\partial x} = 0 \quad ; \quad \frac{\partial h_c}{\partial t} + \frac{\partial}{\partial x}(h_c u_c) = 0 \qquad (3)$$

where the variable channel heights are given by $h_1(x,t) = H_1(x) - y(x,t)$ and $h_2(x,t) = H_2(x) + y(x,t)$. Following previous work [1, 2], localized dissipative effects are enforced at the boundary conditions at $x = 0$ and $x = 1$:

$$p_c(0,t) = P_0 - \frac{1}{2}[u_c(0,t)]^2 - \frac{1}{2}|u_c(0,t)|u_c(0,t)K_0 \;\; ; \;\; p_c(1,t) = P_L - \frac{1}{2}[u_c(1,t)]^2 + \frac{1}{2}|u_c(1,t)|u_c(1,t)K_L \qquad (4)$$

where $K_0$ and $K_L$ are the entry and exit head-loss coefficients, while $P_0$ and $P_L$ are pressures imposed at the entry and exit of the domain. The dynamics of a similar system, considering constant cross-sections, $H_1(x) = H_2(x) = H_0$, were studied recently by the authors [3]. In this short extension, we explore on the effects of a variable channel confinement. More specifically, we will consider symmetric channels that either converge or diverge monotonically along the domain, i.e. a tapered passage. The profile of the confinement walls is then defined by

$$H_c(x) = H^*[1 + 2\beta(x - 0.5)] / 2 \quad \text{with} \quad -1 < \beta < 1 \qquad (5)$$

where $H^* = H_0 / L$ is the confinement ratio and $\beta$ is a slope parameter. Note that $\beta > 0$ represents an expanding passage while $\beta < 0$ denotes narrowing passage. With this definition of channel profile, the average channel height $H_c(x = 0.5) = H_0 / L$ is independent of $\beta$. Hence, relevant dimensionless variables are normalized by this value. Notably, assuming as usual the loss coefficients $K_0 = 0$ and $K_L = 1$, the reduced flow velocity $U^*$ is retrieved from the associated steady problem leading to

$$U^* = 2\pi(1 + \beta)\sqrt{2(P_0 - P_L)} \qquad (6)$$

### Model discretization and continuation methods

For compactness, here we restrain from showing the details of the spatial discretization procedure used to convert the PDE system (3)-(4) into a set of time-dependent equations (ODE/DAE system). The interested reader is referred to the authors previous work [4]. Essentially, the pressure and velocity fields in each channel are developed in terms of a set of orthogonal basis functions

$$u_c(x,t) = \sum_{n=0}^{N} T_n(x) u_n(t) \quad ; \quad p_c(x,t) = \sum_{r=0}^{N} T_r(x) p_r(t) \qquad (7)$$

where $T_n(x)$ are, for example, Chebyshev polynomials of the first kind. After a modified Galerkin projection (Tau-method) on the fluid equations of each channel $c = 1, 2$, and assembly with the structural equations (1)-(2), the resulting coupled system consists of a set of first-order nonlinear differential-algebraic equations (DAE) of size $2M + 4(N + 1)$. Contrary to previously derived formulations [5, 6], the Galerkin approach presented here allows us to discretize the continuous 1-D problem into a set of nonlinear time-dependent equations, compatible for use in algorithms for the continuation of periodic solutions. In this work we have used the open-source software MANLAB 4.0 [7]. which combines the Harmonic Balance Method for the time-discretization (Fourier decomposition) with the Asymptotic Numerical Method for the numerical continuation of the solution path.

### LINEAR STABILITY ANALYSIS

We start by analysis the dynamics of the system in terms of its linear stability. We use the "augmented" linear stability analysis method proposed in [3], which allows the distinction between super- and sub-critical Hopf points and estimate regions of hysteric behaviour. In all results, we fix the confinement ratio $H^* = 0.1$ and beam damping ratio $\zeta = 0.5\%$, and explore the effects of the reduced velocity $U^*$, fluid-beam mass ratio $M^*$ and channel slope parameter $\beta$. In a uniform channel, cantilever beams develop flutter instabilities as a product of the coupling between the passing flow and the first two beam modes [8], i.e. flutter instabilities do not appear when the system is truncated to a single mode. However, our results show that in an expanding passage with sufficient slope, single-mode flutter can also occur. Critical stability maps for a system truncated to a single mode are shown in Figure 2. Note that, as the slope of the expanding passage decreases, critical boundaries are shifted towards very large flow velocities and/or low mass-ratios, until they eventually vanish. These results are in agreement with the work by Inada and Hayama [1, 2], where the stability of a rigid plate in a tapered passage with translational and/or rotational motions (1-DoF and 2-DoF) is studied. Similarly, the authors conclude that diverging passages can generate flutter in 1-DoF systems, while in converging passages typically require two-modes to generate a flutter instability.

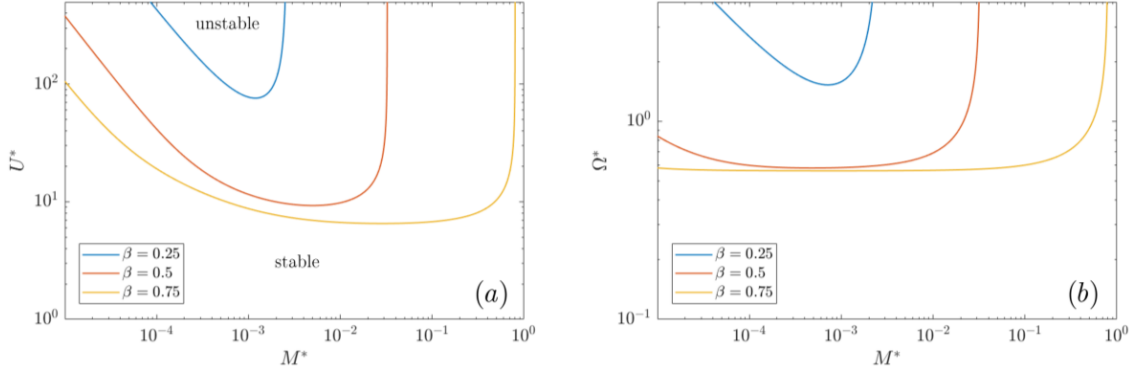

**FIGURE 2**. (a) Critical stability boundaries in the $M^* - U^*$ plane and (b) corresponding instability frequencies $\Omega^*$, for a system with expanding channels $(\beta > 0)$, truncated to a single beam mode $M = 1$, $H^* = 0.1$ and $\zeta_m = 0.5\%$.

In Figure 3, we show the critical boundaries of a system truncated to two beam modes in both expanding $(\beta > 0)$ and narrowing $(\beta < 0)$ passages. Here, we note that Hopf bifurcations (flutter instability) occur for both expanding and narrowing channels, even though in the expanding case, we have two distinct regions, separating single- from coupled-mode flutter (this is clarified in Figure 4(c)). This distinction is also clear from the instability frequencies – around $\Omega^* \approx 1$ for the former and $\Omega^* \approx 3$ for the latter. We note as well that, for the 2-mode system, a narrowing passage tends to lower critical velocities (i.e. destabilises the system) and shift critical boundaries towards lower mass-ratios. Finally, we underline how, contrary to the single-mode flutter in Figure 2 (where all Hopf points are supercritical), here the Hopf bifurcations can be either super or sub-critical, indicating that certain parametric configurations will lead to hysteretic behaviour [9].

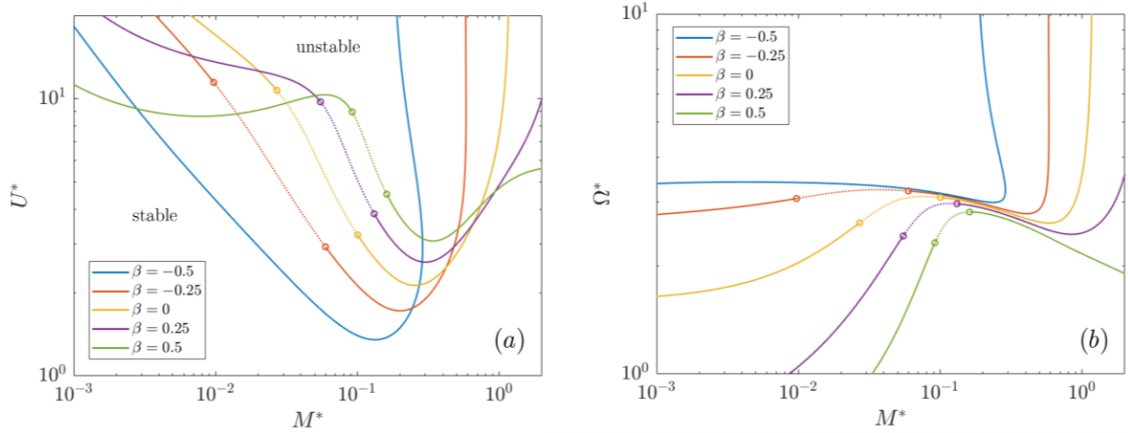

**FIGURE 3**. (a) Critical stability boundaries in the $M^* - U^*$ plane and (b) corresponding instability frequencies $\Omega^*$, for a system with two beam modes $M = 2$, $H^* = 0.1$ and $\zeta_m = 0.5\%$, at various values of channel slope $\beta$. Solid and dotted lines denote super- and sub-critical Hopf bifurcations, respectively, while circles denote Bautin bifurcations.

## NONLINEAR BIFURCATION ANALYSIS

Using the continuation methods mentioned above, branches of periodic solutions stemming from the Hopf (linear stability) boundaries were calculated at constant mass-ratios $M^*$. Once again, the beam motion was truncated to two modes and the temporal discretization was truncated to 15 harmonics $(H = 15)$ Results for two cases of either expanding $(\beta = 0.5)$ or narrowing $(\beta = -0.25)$ profiles are shown in Figure 4, in terms of the maximum tip displacement $A^*$.

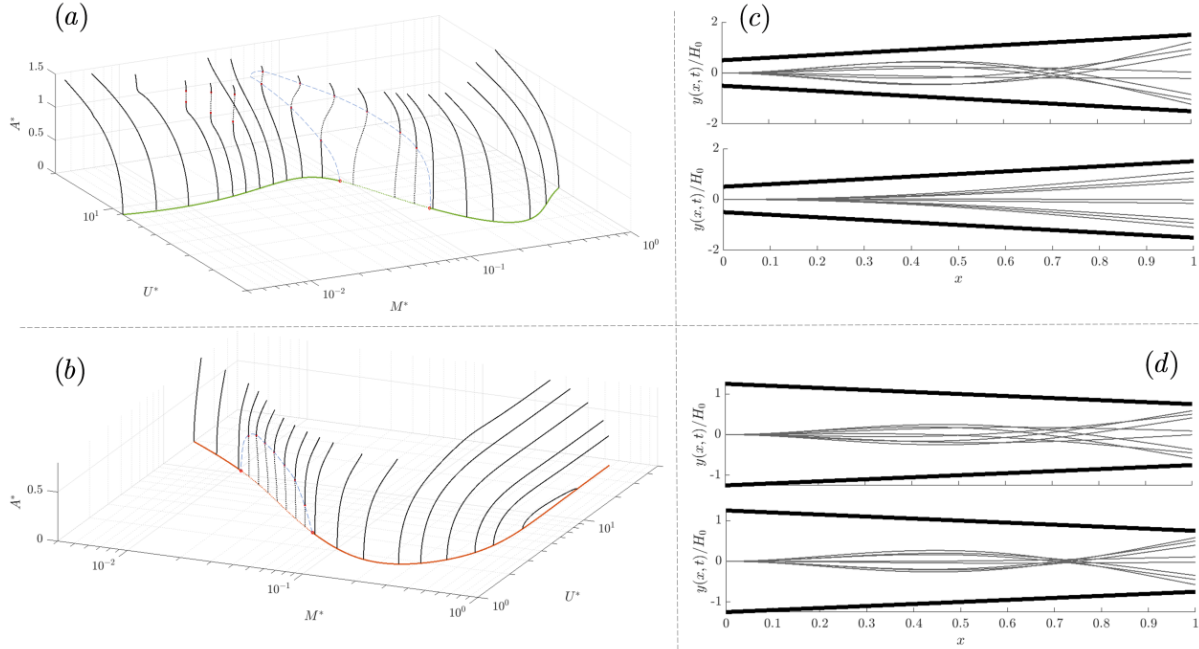

**FIGURE 4**. Plots (a) and (b) are 3D bifurcation diagrams for a system with expanding $(\beta = 0.5)$ and narrowing $(\beta = -0.25)$ passages, respectively. Each black line leaving the Hopf boundary corresponds to a branch of periodic solutions at constant mass-ratio. The dashed blue lines correspond to an estimated Fold bifurcation branch, stemming from the Bautin points mentioned previously. Plots (c) and (d) show snapshots of the beam motion for both cases at two different configurations: $M^* = 0.01$ and $U^* = 12$ (bottom); $M^* = 0.3$ and $U^* = 4$ (top).

In both expanding and narrowing cases we see at least one region where hysteresis loops will take place (delimited by the fold bifurcations – dashed blue line). In the expanding cases however, this region takes a more irregular form. This is likely because, in this case, the hysteretic region is placed near the transition between single- and coupled-mode flutter (see Figure 4(c)) whilst in the narrowing case, the flutter instability is always of the coupled-mode type. Note also that solution branches stop at a certain threshold amplitude, whereby contact with the channel walls occurs. Solution branches in the expanding case are, in general, longer because the beam-tip is allowed larger amplitude motions. Curiously, under the current model simplifications, periodic solutions at large mass-ratios in the narrowing case do not encounter a grazing boundary (wall contact) and the beam motion remains contactless and regular even at very large flow velocities.

## CONCLUSIONS

The nonlinear dynamics of a cantilever beam subject to axial flow in a tapered passage were studied using a one-dimensional bulk-flow model and methods for the continuation of periodic solutions. Contrary to the classical case with parallel walls, our results show that flutter instabilities involving a single beam mode can occur in expanding passages. Otherwise, when two beam modes are considered, a narrowing passage will tend to destabilise the system. The continuation methods used provide a comprehensive view of the system dynamics and show, for example, how hysteresis loops are formed in certain parametric configurations. We hope to have showcased how these methods provide an efficient tool to analyze the nonlinear behavior of flow-induced vibration problems, particularly when reduced models are available.

## REFERENCES

1. Inada & Hayama, A study on leakage-flow-induced vibrations. Part 1: Fluid-dynamic forces and moments acting on the walls of a narrow tapered passage, J. Fluids Struct, 1990.
2. Inada & Hayama, A study on leakage-flow-induced vibrations. Part 2: Stability analysis and experiments for two-degree-of-freedom systems combining translational and rotational motions, J. Fluids Struct, 1990.
3. Soares et al., Bifurcation analysis of a cantilever beams in channel flow, J. Sound Vib, 2023.
4. Soares et al., A Galerkin formulation for the nonlinear analysis of a flexible beam in channel flow, J. Fluids Struct, 2023.
5. Nagakura & Kaneko, The stability of a cantilever beam subjected to one-dimensional leakage flow. SMiRT 11, 1991.
6. Soares et al., A nonlinear analytical formulation for the 1D modelling of a flexible beam in channel flow, J. Fluids Struct, 2022.
7. Manlab: An interactive path-following and bifurcation analysis software, [Online: http://manlab.lma.cnrs-mrs.fr/].
8. Auregan Y., Contribution à l'étude des bruits respiratoires: modélisation du ronflement, PhD Thesis, Université du Maine, 1993.
9. Eloy et al., The origin of hysteresis in the flag instability, J. Fluid Mech, 2011.